\documentclass[AMA,STIX1COL]{WileyNJD-v2}

\usepackage{moreverb,url}

\usepackage{mathrsfs,bm}
\usepackage{dsfont}
\usepackage{float}
\usepackage{bbm}
\usepackage{amsmath}
\usepackage{graphicx,subfig}
\usepackage{amsthm}
\usepackage{verbatim}
\usepackage{epsfig,epstopdf}
\usepackage[labelfont=bf]{caption}
\usepackage{enumerate}
\usepackage{epstopdf}
\usepackage{tikz,circuitikz}
\usepackage{xcolor}

\makeatletter
\newcommand{\pushright}[1]{\ifmeasuring@#1\else\omit\hfill$\displaystyle#1$\fi\ignorespaces}
\makeatother

\def\begquo{\begin{quote}}
	\def\endquo{\end{quote}}
\def\begequarr{\begin{eqnarray}}
	\def\endequarr{\end{eqnarray}}
\def\begequarrs{\begin{eqnarray*}}
	\def\endequarrs{\end{eqnarray*}}
\def\begarr{\begin{array}}
	\def\endarr{\end{array}}
\def\begequ{\begin{equation}}
	\def\endequ{\end{equation}}
\def\lab{\label}
\def\begdes{\begin{description}}
	\def\enddes{\end{description}}
\def\begenu{\begin{enumerate}}
	\def\begite{\begin{itemize}}
		\def\endite{\end{itemize}}
	\def\endenu{\end{enumerate}}

\def\lef[{\left[\begin{array}}
	\def\rig]{\end{array}\right]}
\def\qed{\hfill$\Box \Box \Box$}
\def\begcen{\begin{center}}
	\def\endcen{\end{center}}
\def\begrem{\begin{remark}\rm}
	\def\endrem{\end{remark}}
\def\begdef{\begin{definition}}
	\def\enddef{\end{definition}}
\def\begpro{\begin{proposition}}
	\def\endpro{\end{proposition}}
\def\begfac{\begin{fact}}
	\def\endfac{\end{fact}}
\def\begass{\begin{assumption}}
	\def\endass{\end{assumption}}

\def\begsubequ{\begin{subequations}}
	\def\endsubequ{\end{subequations}}

\def\begmat#1{\begin{bmatrix}#1\end{bmatrix}}
\def\begali#1{\begin{align}{#1}\end{align}}
\def\begalis#1{\begin{align*}{#1}\end{align*}}

\def\calg{{\cal G}}

\def\calh{{\cal H}}
\def\calc{{\cal C}}

\def\calb{{\cal B}}

\def\calw{{\cal W}}

\def\cala{{\cal A}}

\def\caly{{\cal Y}}

\def\liminf{\lim_{t \to \infty}}

\def\L2e{{\cal L}_{2e}}

\def\rea{\mathds{R}}
\def\intnum{\mathds{N}}

\def\diag{\mbox{diag}}
\def\adj{\mbox{adj}}
\def\col{\mbox{col}}
\def\hal{{1 \over 2}}

\def\diag{\mbox{diag}}
\def\rank{\mbox{rank}\;}

\def\SIAM{{\it SIAM J. Control and Optimization}}
\def\IJACSP{{\it Int. J. on Adaptive Control and Signal Processing}}
\def\IJRNLC{{\it Int. J. on Robust and Nonlinear Control}}
\def\TAC{{\it IEEE Trans. Automatic Control}}

\def\TPE{{\it IEEE Trans. Power Electronics}}

\def\IJC{{\it International Journal of Control}}

\def\SCL{{\it Systems and Control Letters}}
\def\AUT{{\it Automatica}}

\def\CSM{{\it IEEE Control Systems Magazine}}

\def\calg{{\cal G}}

\def\hal{{1 \over 2}}

\usepackage{color}

\def\bfe{{\bf e}}
\def\begsubequ{\begin{subequations}}
	\def\endsubequ{\end{subequations}}
\def\begpro{\begin{proposition}}
	\def\endpro{\end{proposition}}
\def\beglem{\begin{lemma}}
	\def\endlem{\end{lemma}}
\def\begass{\begin{assumption}}
	\def\endass{\end{assumption}}
\def\begcor{\begin{corollary}}
	\def\endcor{\end{corollary}}
\def\begproo{\begin{proof}}
	\def\endproo{\end{proof}}

\usepackage[prependcaption,colorinlistoftodos]{todonotes}

\articletype{Article Type}%

\received{26 april 2016}
\revised{6 june 2016}
\accepted{6 june 2016}
\begin{document}

\title{Immersion of General Nonlinear Systems Into State-Affine Ones for the Design of Generalized Parameter Estimation-Based Observers: A Simple Algebraic Procedure}

\author[1]{Romeo Ortega}
\author[2]{Alexey Bobtsov}
\author[1]{Jose Guadalupe Romero*}
\author[1,3]{Leyan Fang}

\authormark{R. Ortega and et al;}

\address[1]{\orgdiv{Departamento Acad\'{e}mico de Ingenier\'ia El\'ectrica y Electr\'onica}, \orgname{ITAM}, \orgaddress{\state{Ciudad de M\'exico}, \country{M\'exico}}}

\address[2]{\orgdiv{Control Systems and Robotics Department}, \orgname{ITMO}, \orgaddress{\state{St. Petersburg}, \country{Rusia}}}

\address[3]{\orgdiv{Center for Control Theory and Guidance Technology}, \orgname{ Harbin Institute of Technology}, \orgaddress{\state{Harbin}, \country{China}}}

\corres{*J. G. Romero, Departamento Acad\'{e}mico de Ingenier\'ia El\'ectrica y Electr\'onica, ITAM, R\'io Hondo 1, 01080,  Ciudad de M\'exico, M\'{e}xico. \email{jose.romerovelazquez@itam.mx}}

\abstract[Summary]{
 Generalized parameter estimation-based observers have proven very successful to deal with systems described in state-affine form. In this paper, we enlarge the domain of applicability of this method proposing an {\it algebraic} procedure to {\it immerse} an $n$-dimensional general nonlinear system into and $n_z$-dimensional system in state affine form, with $n_z>n$. First, we recall the {\it necessary and sufficient} condition for the solution of the general problem, which requires the solution of a partial differential equation that, moreover, has to satisfy a restrictive {\it injectivity} condition. Given the complexity of this task we propose an alternative  simple {\it algebraic method} to identify the required dynamic extension and coordinate transformation, a procedure that, as shown in the paper, is rather natural for physical systems. We illustrate the method with some academic benchmark examples from observer theory literature---that, in spite of their apparent simplicity, are difficult to solve with the existing methods---as well as several practically relevant physical examples.   
}


\maketitle

%
\section{Introduction}
\lab{sec1}
%

A class of systems for which the problem of designing a state observer is well-known, are the so-called {\it state-affine} systems, whose dynamics is described by \cite[Equation 3.1]{WANORTBOB}:
\begali{
\nonumber
\dot x &=\cala(y,u)x+\calb(y,u)\\
\lab{staaff}
y & = h(x,u),
}
where $x(t) \in \rea^n$ is the {\it unmeasurable} state, $y(t)  \in \rea^p$ denotes the measured output signal and $u(t) \in \rea^m$ is the control input. When the read-out map is linear in $x$, that is, $h(x,u)=\calc(u)x$ the most famous observer used for this kind of systems is the Kalman and Bucy's observer introduced in \cite{KALBUC} for linear time-varying (LTV) systems---which is obtained evaluating along trajectories the matrices $\cala,\calb$  and $\calc$ of the system above. It is well-known that to ensure global convergence for this observer it is necessary to impose very strong excitation assumptions on the system, namely, {\it uniform complete observability} of the pair $(\cala(y(t),u(t)),\calc(u(t)))$ \citep{BERbook, KALBUC}. In a recent paper \cite{WANORTBOB} it was shown that state observation of LTV systems is possible imposing only the (necessary) assumption of {\it observability}. For, it is proposed to use the recently reported {\it generalized parameter estimation-based observers} (GPEBO) \cite{ORTetalaut}. The main feature of GPEBO is that the problem of state observation is recasted as a problem of {\it parameter} estimation, namely of the systems initial conditions. This approach has proven to be very successful for the state observation of state-affine systems, and many extensions and practical applications to the method have been reported \cite{BEZetal,BOBetalaut21,BOBetalijc,ORTetaltac,PYRetal,ROMORT}.

In this paper, we enlarge the domain of applicability of GPEBO proposing a procedure to {\it immerse} an $n$-dimensional {\it general nonlinear system} into and $n_z$-dimensional state affine system of the form \eqref{staaff}, with $n_z>n$.\footnote{Notice that we do not assume that $h(x,u)=C(u)x$.} The problem of designing a change of coordinates to transform a general nonlinear system into a particular form suitable for observer design has a very long history, dating from the pioneering work of \cite{KREISI}. Several extensions to this result have been reported over the years, culminating with the interesting result  where they observed that it is not necessary to linearize the output map, but simply obtain a transformed dynamics which is linear and is driven by a nonlinear output injection\cite{KAZKRA}---see the recent tutorial paper \cite{BERANDAST} for a historic account of this line of research and an insightful description of the difficulties encountered in this method. The first time that instead of looking for a diffeomorphism that transforms the system into observer form, motivated by \cite{FLIKUP}, it was investigated the possibility of {\it immersing} it into a higher dimensional observer form was reported in \cite{LEVMAR}. This interesting approach has also been pursued by \cite{BACYUSEO,BACSEO}. 

All these methods described above suffer from the limitation that it is necessary to solve  a {\it partial differential equation} (PDE) and ensure that the resulting diffeomorphism is {\it injective}.  To overcome these difficulties we propose in this paper to adopt an {\it algebraic} approach to select  the system immersion and required coordinate transformation. It is shown in the paper that this procedure is quite systematic and rather natural for physical systems. 

The transformation of the system into a state-affine one allows us to fulfill the first step in the design of GPEBO, which proceeds along the following three steps.

\begenu[{\bf S1}]
\item The expression of the unmeasurable state as a sum of measurable signals and an {\it unknown} parameter vector, related with the states initial conditions.
\item Derivation of a {\it regression equation} to implement the parameter estimation, which depending on the form of the read-out map may be a linear regression equation (LRE) or a nonlinearly parameterized regression equation (NLPRE). Further, in the case when the read-out map contains trascendental functions, we are in the worst scenario of {\it non-separable} regression equations for which only a few results are available  \cite{NEVMIRMBO,ORTetalaut24,USHetal}.
\item Characterization of the excitation conditions needed for parameter convergence---the weakest one been {\it interval excitation} (IE) \cite{KRERIE}. Actually, in \cite{WANORTBOB} it is proven that IE is a {\it necessary and sufficient} condition for estimation of the parameters of a LRE. 
\endenu 
All these issues are discussed in the examples given in the paper, which contain  some academic benchmark examples from observer theory literature, where the difficulty to design observers for them is discussed. We also apply our observer design method to three practically relevant physical examples, that have been widely studied in the literature: namely, magnetic levitation systems, permanent magnet synchronous motors (PMSM) and a class of mechanical systems.      

The rest of the paper has the following structure. To motivate our algebraic procedure, in Section \ref{sec2} we present a {\it general} formulation of the problem we address in the paper and present the {\it necessary and sufficient} condition for its solution. As discussed above this involves the solution of a PDE that, moroever, has to satisfy a restrictive {\it injectivity} condition, hence it is of little practical interest.  The main result of the paper---that is, a simple constructive, {\it algebraic} immersion procedure---is given in Section \ref{sec3}. In Section \ref{sec4} we present several examples of application  of the method.  In Section \ref{sec5} we present some simulation results. Finally, in Section \ref{sec6} we include some concluding remarks. In Appendix \ref{appa} we recall the Least-Squares + Dynamic Regressor Extension and Mixing (LS+DREM)  parameter estimator\cite{ORTROMARA}, which is used in the simulations and enjoys two key features: that parameter convergence is ensured imposing the extremely weak {\it interval excitation} (IE) assumption and that it can handle NLPRE.\\ 

\noindent {\bf Notation.} $I_n$ is the $n \times n$ identity matrix.  ${0}_{s \times r}$ is an $s \times r$ matrix of zeros. $\intnum_{>0}$ denotes the positive integer numbers. $\col(x_1,\cdots,x_n)$ denotes the column vector $x \in \rea^n$. The vector $\bfe_j \in \rea^n$, with $j \in\{1,2,\cdots,n\}$, is the $j$-th element of the unitary Euclidean $n$-dimensional basis. For a differentiable mapping $V:\rea^n \to \rea$ we define $\nabla_{x_i}V(x):={\partial V(x) \over \partial x_i}$ and the gradient transpose operator $\nabla V(x):=\Big({\partial V(x) \over \partial x}\Big)^\top$.   
\section{The Mathematically Intractable General Problem Formulation}
\lab{sec2}
%
In this section we present a {\it general formulation} of the state transformation plus immersion problem that we address in the paper. We also give the well-known {\it necessary and sufficient} condition for its solvability---namely the {\it solution of a PDE} that, moreover, has to satisfy a restrictive {\it injectivity} condition. Our objective in presenting this general formulation is to underscore the mathematical intractability of this task that provides motivation for our {\it partially constructive} algebraic procedure presented in the following section. 
\subsection{A change of coordinates and systems immersion problem}
\lab{subsec21}
%
Consider a general nonlinear system of the form 
\begsubequ
		\lab{sys}
		\begali{
			\lab{sysdotx}
			\dot x=&\, f(x,u)\\
	\lab{sysy}
		y=&\, h(x,u),
}
\endsubequ
where $x(t) \in \rea^n$ is the {\it unmeasurable} state, $y(t)  \in \rea^p$ denotes the measured output signal and $u(t) \in \rea^m$ is the control input. Give conditions for the existence of a dynamical system
		\begali{
			\lab{dotw}
			\dot w=&\, r(w,y,u)
}
with $w(t) \in \rea^{n_w}$ and a diffeomorphism $T_w(x,w)$ that transforms---via $z_w=T_w(x,w)$---the augmented system
		\begali{
			\lab{augsys}
			\begmat{  \dot x \\ \dot w } &=\begmat{ f(x,u) \\ r(w,h(x,u),u)\\},
}
into an $n+n_w$-dimensional {\it state-affine} system of the form
		\begali{
			\lab{augsysaff}
			\dot z_w&=\cala_w(w,y,u)z_w+\calb_w(w,y,u),
}
with $\cala_w:\rea^{n_w} \times \rea^p \times \rea^m \to \rea^{(n+n_w) \times (n+n_w)}$ and $\calb_w:\rea^{n_w} \times \rea^p \times \rea^m \to \rea^{n+n_w}$.

\begrem
\lab{rem1}
See \cite[Theorem 1]{BACYUSEO} for the case of transformation to the classical {\it observable} linear system form. 
\endrem
\subsection{A necessary and sufficient condition}
\lab{subsec22}
%
The proposition below identifies the well-known necessary and sufficient condition for the solution of the problem posed above. 
\begin{proposition}
		\lab{pro1}
There exists a change of coordinates $z_w=T_w(x,w)$ that transform the augmented system \eqref{augsys} into \eqref{augsysaff} {\it if and only if} the mapping $T_w(x,w)$ satisfies the PDE
\begequ
\lab{pdeaug}
\nabla^\top T_w(x,w)\begmat{ f(x,u)\\r(w,h(x,u),u)}=\cala_w(w,h(x,u),u)T_w(x,w) +\calb_w(w,h(x,u),u).
\endequ
\qed		
\end{proposition}
In this case the GPEBO design would then proceed introducing the dynamic extension
\begalis{
\dot \xi &=\cala_w(w(t),y(t),u(t))\xi+\calb_w(w(t),y(t),u(t)),\;\xi(0)=\xi_0 \in \rea^{n+n_w}\\
\dot \Phi &= \cala_w(w(t),y(t),u(t)) \Phi,\;\Phi(0)=I_{n+n_w}.
}
Following the standard GPEBO procedure \cite{ORTetalaut} we derive a parameterization for the unknown state $z_w$ as
$$
z_w=\xi+\Phi \theta,
$$
where $\theta \in \rea^{n+n_w}$ is an {\it unknown} parameter vector, which is estimated using the NLPRE
$$
y=h(T_w^I(\xi+\Phi \theta,w),u),
$$
with the mapping $T_w^I:\rea^{n+n_w}\times \rea^{n_w} \to \rea^n$ verifying
\begequ
\lab{tw}
T_w^I(T_w(x,w),w)=x,\;\forall w(t) \in \rea^{n_w}.
\endequ
\begrem
\lab{rem2}
It is important to underscore that we leave aside the important issue of ensuring that $T_w$ is {\it injective}---that is, that it satisfies \eqref{tw} for some map $T_w^I$---to be able to recover the original state $x$ from $(z_w,w)$. See \cite{ANDPRA,BERANDAST} for a thorough discussion of this issue in the context of Kazantzis-Kravaris observers \cite{KAZKRA} and \cite{BERbook} for the general observer case.
\endrem
\section{A Simple Algebraic Immersion Procedure}
\lab{sec3}
%
To overcome the difficulties encountered in the solution of the general problem above, in this section we propose a simple {\it algebraic} procedure to immerse a general nonlinear system into a state-affine one, making it suitable for the design of a GPEBO. 
\subsection{Main result}
\lab{subsec31}
%
\begin{proposition}
\lab{pro2}
Consider the general nonlinear system \eqref{sys}. Assume there exist mappings
\begalis{
& W_{1,1}:\rea^p \times \rea^m \to \rea^{n \times n},\;W_{1,2}:\rea^p \times \rea^m \to \rea^{n \times \ell},\;L_1:\rea^p \times \rea^m \to \rea^{n} \\
& W_{2,1}:\rea^p \times \rea^m \to \rea^{\ell \times n},\;W_{2,2}:\rea^p \times \rea^m \to \rea^{\ell \times \ell},\;L_2:\rea^p \times \rea^m \to \rea^{\ell},
}
and $\phi:\rea^n \to \rea^{\ell}$ verifying the following equations.
\begsubequ
\lab{matequ}
\begali{
\lab{matequ1}
f(x,u)&=  W_{1,1}(h(x,u),u)x + W_{1,2}(h(x,u),u)\phi(x)+  L_{1}(h(x,u),u) \\
\lab{matequ2}
\nabla^\top \phi(x) f(x) &= W_{2,1}(h(x,u),u)x + W_{2,2}(h(x,u),u)\phi(x)+  L_{2}(h(x,u),u).
}
\endsubequ
Then, the following relation holds
\begali{
\lab{dotz}
\dot z &= W(y,u)z+L(y,u),
}
where we defined the vector\footnote{That is, $\phi(x)=\col(z_{n+1}, \cdots, z_{n_z})$.}
$$
z:=\begmat{x \\ \phi(x)} \in \rea^{n_z},\;n_z:=n+\ell
$$
and the matrices
$$
 W(y,u):=\begmat{ W_{1,1}(y,u) & W_{1,2}(y,u) \\ W_{2,1}(y,u) & W_{2,2}(y,u)},\;L(y,u):=\begmat{  L_{1}(y,u) \\ L_{2}(y,u)}.
$$
\end{proposition}
\begproo
The proof is easily established noting that
$$
\dot z=\begmat{f(x) \\ \nabla^\top \phi(x) f(x)}
$$
and replacing the matching equations \eqref{matequ}. 
\endproo
Under the conditions of Proposition \ref{pro2}, the first step in the GPEBO design can be seamlessly realized. Namely, defining the dynamic extension
\begali{
\nonumber
\dot \xi&=W(t) \xi+ L(t)\\
\lab{gpebo}
\dot \Phi&=W(t) \Phi,\;\Phi(0)=I_{n_z},
}
where we defined
\begalis{
W(t)&:=W(y(t),u(t))\\
L(t)&:= L(y(t),u(t)),
}
we obtain the the well-known parameterization 
$$
z=\xi-\Phi \theta,
$$
with $\theta:=\xi(0)-z(0)$ an {\it unknown} parameter vector. Using the fact that, with $D:=\begmat{I_n  & \vdots & 0_{n \times \ell} }$, we have that $x=Dz$, hence we obtain the following parameterization of the unknown state $x$
\begequ
\lab{x}
x=D\xi-D\Phi\theta.
\endequ
\subsection{Discussion}
\lab{subsec32}
%
\begenu[{\bf D1}]
\item To solve the matching equations \eqref{matequ} we proceed as follows. First, identify the elements of the vector $f(x,u)$ which contain terms that depend nonlinearly on components of the vector $x$ that cannot be expressed as functions of $h(x,u)$---for instance, if $y=x_1$ and there is a term $x_2^2$. Include this term as an element of the vector $\phi(x)$, say $\phi_j(x)=x^2$. The condition \eqref{matequ2} imposes a constraint on the derivative of $\phi_j(x)$, capturing the fact that it should be possible to express it as an affine combination of $x$ and $\phi(x)$, with weighting factors functions of $y$. This is, in essence, the constraint that is imposed for the application of the method, which is illustrated in the examples of Section \ref{sec4}. Notice that, adopting this viewpoint,  \eqref{matequ2} {\it should not be viewed} as a PDE on $\phi(x)$, but as an algebraic constraint.
\item As discussed in step {\bf S2} of the GPEBO design explained in the Introduction to carry out the estimation of the parameters stemming from the GPEBO technique it is necessary to develop a {\it regression equation}. In the present case where no assumption is made on the readout map $h(x,u)$ in \eqref{sysy} this is a NLPRE and is obtained replacing \eqref{x} in the read-out map to get
\begequ
\lab{yequh}
y=h(D\xi-D\Phi\theta,u).
\endequ
On the other hand, if the read-out map $h(x,u)$ is {\it linear in $x$}, that is, $h(x,u)=C(u)x$, with $C:\rea^m \to \rea^{p \times n}$ then we get a LRE
\begequ
\lab{lreypsi}
\caly=\psi\theta,
\endequ
where we defined the measurable signals
\begali{
\nonumber
\caly & :=y-C(t)\xi\\
\lab{psi}
\psi &:=C(t) \Phi
}
with $C(t):=C(u(t))$.
\item Notice that, if  $h(x,u)$ does not contain {\it trascendental functions in $x$} in it is easy to show that, after some simple manipulations, we can write \eqref{yequh} in the {\it separable} form
\begequ
\lab{sepregegu}
\caly=\psi\calg(\theta),
\endequ
where $\caly(t) \in \rea^p$ and $\psi(t) \in \rea^{p \times n_\psi}$ are {\it measurable} and $\calg:\rea^{n_z} \to \rea^{n_\psi}$, where $n_\psi \geq n_z$. As is well-known, it is (sometimes) possible to deal with the problem of NLPRE using a modern parameter estimator, {\it e.g.}, the LS+DREM estimator\cite{ORTROMARA}---see also Appendix \ref{appa}. If, on the other hand, there are trascendental functions in $h(x,u)$, then we are dealing with a much more complicated situation, since in this case the regression equation is {\it not separable}, that is, it is of the form
\begequ
\lab{nonsep}
\caly(t)=\varphi(t,\theta),
\endequ
where $\varphi: \rea_+ \times \rea^{n} \to \rea^r$. Some preliminary results for this case, specifically for the often encountered case of exponential or cosine functions, may be found in \cite{NEVMIRMBO,ORTetalaut24,USHetal}. See the example of Subsection \ref{subsec44}.
\item The third step of the GPEBO design, namely, the assessment of the {\it excitation} requirements to ensure parameter convergence of the LS+DREM estimator is a condition imposed on the regressor vector $\psi$. For LRE or NLPRE it is simply that it is IE \cite{KRERIE}: namely, that there exists $t_c>0$ and $\delta>0$ such that
$$
\int_0^{t_c} \psi^\top(s)\psi(s)ds > \delta I_{n_\psi}.
$$
Replacing the definition of the regressor for the case of linear read-out map given in \eqref{psi} we get
$$
\calw_O(0,t_c):=\int_0^{t_c}  \Phi^\top(s) C^\top(s) C(s) \Phi(s) ds > \delta I_{n_\psi},
$$
where $\calw_O(0,t_c)$ is the {\it observability Grammian} of the LTV system $(W(t),C(t))$ \cite[Definition 15.5]{HESbook}. This fact proves the fundamental result that 
$$
\psi(t)\in IE\quad \Leftrightarrow \quad (W(t),C(t)) \mbox{\;is\;observable}.
$$
We recall also that in \cite[Lemma 3]{WANORTBOB} it is shown that---for the single output LRE case---IE is {\it equivalent} to the existence of a time sequence $\{t_j\}_{j=1}^{n_z}$ such that 
$$
\rank\Bigg\{\begmat{\psi(t_1)\\\vdots\\\psi(t_{n_z})}\Bigg\}=n_z,
$$
that is the definition of (off- or on-line) {\it identifiability} of the LRE \eqref{lreypsi} \cite{GOOSINbook}. From the previous analysis we conclude that GPEBO ensures convergence imposing only the {\it necessary} assumption of observability of the system (equivalently, identifiability of the LRE).
\item Throughout the paper no reference is made to the issue of the stability of the GPEBO design---in particular, the {\it global boundedness} of all signals. It is clear from \eqref{gpebo} that this is determined by the matrix $W(y(t),u(t))$. If the associated LTV system is {\it unstable}, the trajectories of the GPEBO will be unbounded---this may happen even if (as it is customarily assumed) the systems state trajectories are bounded. This issue has been discussed in \cite{BOBetalaut21} and, as explained there, the problem can be (partially) overcome implementing an estimator with finite convergence time. 
\item It is possible to consider the case when the dynamical system is described by
$$
\dot x=\, f(x,u)+d(h(x,u),u)\eta,
$$
where $\eta \in \rea^{n_\eta}$ is an {\it unknown} constant vector---in these case we can easily implement an adaptive GPEBO\cite{ORTetalaut}, if the parameters are {\it time-varying} we should follow \cite{BOBetalijc}. Similarly, it is possible to consider the case when the output signal is measured with a {\it delay} or the state equation is perturbed by an exogenous signal, whose internal model is {\it unknown}. To address these two issues the interested reader is referred to  \cite{BOBetalaut21} and 
 \cite{PYRetal}, respectively.
\endenu
\section{Benchmark Examples}
\lab{sec4}
%
In this section we work out several examples including some academic ones reported in the observer theory literature.  It should be underscored that, even though the academic examples look quite simple, as indicated below it  has been reported in the literature they {\it they are not} amenable for their solution using the existing observer design tools or they require complicated {\it ad hoc} modifications. We also design our observer for three physical systems that have been extensively studied in the observer design literature: namely, a magnetic levitation system, a PMSM and a class of mechanical systems.
\subsection{An academic example \cite{LEVMAR}}
\lab{subsec41}
%
In this subsection we consider an example reported in  \cite{LEVMAR} that---as shown in  \cite{BACSEO}---requires a dynamic extension to transform it to observer form.

Consider the system 
\begalis{
\dot x &= \begmat{x_2+\hal x_2^2+\alpha(x_1) \\ x_2},\;y =x_1,
}
with $\alpha:\rea \to \rea$. This system is (locally) observable but, as shown in \cite{LEVMAR}, {\it cannot be transformed into observer form} with a standard change of coordinates.\footnote{On the other hand, it is shown in \cite{BACSEO} that the system is immersible into a three-dimensional observer form. See also the interesting paper \cite{BACYUSEO}.}

We observe that this example fits in the scenario discussed in {\bf D1}. Therefore, we select $\phi(x)$ proportional to $x_2^2$, say $\phi(x_2) =\hal x_2^2$. The matching condition \eqref{matequ2} imposes a constraint on 
$$
\dot \phi(x_2)=\phi'(x_2) \dot x_2=x_2^2.
$$
Namely, that it should be expressed as a linear combination of $\phi(x_2)$, which is clearly satisfied with the choice below. Hence, selecting $n_z=3$  and choosing the mappings
\begalis{
W&=\begmat{0 & 1 & 1\\ 0 & {1} & 0 \\ 0 & 0 & 2},\;L(x_1)=\begmat{\alpha(x_1) \\ 0 \\ 0},
}
yields the system \eqref{dotz}. The corresponding GPEBO is given as
\begalis{
\dot \xi&=W\xi+L(y)\\
\dot \Phi&=W \Phi,\;\Phi(0)=I_{3}.
}
and the LRE \eqref{lreypsi} with
\begequ
\lab{lre}
\caly:=\xi_1-y,\; \psi:=\bfe_1^\top\Phi.
\endequ

\begrem
\lab{rem3}
In this example since the original system is unstable the matrix $W$ is non-Hurwitz.
\endrem
\subsection{Two academic examples\cite{BERbook}}
\lab{subsec42}
%
In this subsection we consider two examples reported in  \cite{BERbook}.\\

\noindent{\bf First example:} In  \cite[Example 7.1]{BERbook} the following system is given  
\begalis{
\dot x &= \begmat{x_2\\ x_3^3\\1+u},\;y =x_1.
}
It is shown in \cite{BERbook} that this system {\it does not} admit a classical high-gain design \cite{EMEetal,TOR}, however it is easy to show that it fits into the framework of Proposition \ref{pro1}, hence GPEBO can be easily applied. 

In this example we should select one element of $\phi(x)$, proportional to $x_3^3$, say $\phi_2(x_3)={1 \over 3}x_3^3$. But the matching condition \eqref{matequ2} imposes the constraint 
$$
\dot \phi_2(x_3)=\phi'_2(x_3) \dot x_3=x^2_3(1+u) .
$$
Therefore, we need to introduce and additional element to the vector $\phi(x)$ proportional to $x_3^2$. In summary, we select $n_z=5$ and choose the mappings
\begalis{
\phi(x) &=\begmat{{1 \over 3} x_3^3 \\ \hal x_3^2},\;W(u) =\begmat{0 & 1 & 0 & 0 & 0\\ 0 & 0 & 0 & 3 & 0 \\ 0 & 0 & 0 & 0 & 0 \\ 0 & 0 & 0 & 0 & 2(1+u) \\ 0 & 0 & 1+u & 0 & 0},\;L(u)=\begmat{0\\0\\1+u\\0\\0},
}
yields the  system \eqref{dotz}. The corresponding GPEBO is given as
\begalis{
\dot \xi&=W(u)\xi+L(u)\\
\dot \Phi&=W(u) \Phi,\;\Phi(0)=I_{5}.
}
with the LRE  \eqref{lreypsi}, \eqref{lre}.\\

\noindent{\bf Second example:} The system in  \cite[Example 7.2]{BERbook} is a slight variation of the previous one and given as 
\begalis{
\dot x &= \begmat{x_2\\ x_3^3 x_1\\ 1+u},\;y =x_1.
}
Again, it is shown in \cite{BERbook} that this system {\it does not} admit a classical high-gain design, however a variation of this observer---which is an extremely involved design---is given in \cite[Example 7.3]{BERbook}. 

To prove that it fits into the framework of Proposition \ref{pro1} we select $n_z=5$, $\phi(x)$ as given above, and choose
\begalis{
W(y,u) =\begmat{0 & 1 & 0 & 0 & 0\\ 0 & 0 & 0 & 3y & 0 \\ 0 & 0 & 0 & 0 & 0 \\ 0 & 0 & 0 & 0 & 2(1+u) \\ 0 & 0 & 1+u & 0 & 0},\;L(u)=\begmat{0\\0\\1+u\\0\\0}.
}
This yields the system \eqref{dotz}. The resulting GPEBO  and the LRE are, up to the new $W$, identical to the ones above. 

\begrem
\lab{rem4}
A transformation similar to the one used in the two examples above is proposed in \cite[Subsection 6.3.4]{BERbook} to treat the system 
\begalis{
\dot x &= \begmat{x_2^3\\-x_1},\;y =x_1.
}
As indicated in \cite{BERbook}, writing the system in the coordinates $z=\col(x_1,x_2,x_2^2,x_2^3)$ yields the state-affine dynamics
$$
\dot z = \begmat{0 & 0 & 0 & 1\\-1 & 0 & 0 & 0\\0 & -2y & 0 & 0\\0 & -3y & 0 & 1}z.
$$
As noted in \cite{BERbook} a Kalman-Bucy observer is inappropiate because the required uniform complete observability conditions are extremely restrictive---notice that this is not the case of GPEBO that only requires IE \cite{WANORTBOB}. A variation of this method, considering {\it time-varying} transformations is then proposed, involving a complicated procedure needed to compute the inverse transformation.  
\endrem
\subsection{Magnetic levitation system}
\lab{subsec43}
%
Consider the magnetic levitation system \cite[Example 1.1]{ASTKARORTbook} with state vector the flux $x_1(t) \in \rea$, position $x_2(t) \in(-\infty,1)$ and momenta $x_3(t) \in \rea$, whose dynamics may  be written as
\begalis{
\dot x_1 &= -{R \over k}(1-x_2)x_1+u\\
\dot x_2 &= {1 \over m} x_3\\
\dot x_3 &= {1 \over 2k}x_1^2-mg\\
y &={1 \over k}(1-x_2)x_1,
}
where $u(t)\in \rea$ is the applied voltage, $R,k,m,g$ are positive parameters and the measurable quantity is the {\it current} $y(t) \in \rea$. Notice that the first equation---stemming from Faraday's law---yields the relation
$$
\dot x_1 = -Ry+u.
$$

Again, the disturbing term is $ {1 \over 2k}x_1^2$, therefore we select  $\phi(x_1) ={1 \over 2k} x_1^2$.  It's derivative yields
$$
\dot \phi(x_1)=\phi'(x_1) \dot x_1={1 \over k}x_1(-Ry+u) .
$$
Hence, selecting $n_z=4$ and choosing the mappings
\begalis{
W(y,u) &=\begmat{0 & 0 & 0 & 0\\ 0 & 0 & {1 \over m} & 0\\ 0 & 0 & 0 & 1\\ -\frac{1}{k}(Ry-u) & 0 & 0 & 0},\;L(y,u)=\begmat{-R y + u\\ 0 \\ -mg \\ 0},
}
yields the  system \eqref{dotz}. To get the regression equation we replace \eqref{x} in the output signal $y$. Denoting 
$$
\Phi=\begmat{\Phi^\top_1 \\ \vdots \\  \Phi^\top_4},
$$ 
where $\Phi_i(t) \in \rea^4,\;i=1,\dots,4,$ we get the NLPRE
$$
ky-\xi_1+\xi_1 \xi_2=[(\xi_2-1)\Phi_1 + \xi_1\Phi_2]^\top \theta -\Phi_1^\top \theta \theta^\top\Phi_2.
$$
This equation can be expressed in the separable NLRE form \eqref{sepregegu}, with the following definitions
\begalis{
\caly &:=ky-\xi_1+\xi_1 \xi_2 \\
\psi & := \begmat{[(\xi_2-1)\Phi_1 + \xi_1\Phi_2]^\top& \vdots & - \psi^\top_0}\\  
\calg(\theta)& :=\begmat{\theta \\ \calg_0(\theta)},
}
with 
$$
\psi_0:= \begmat{\Phi_{11} \Phi_{21}\\ \Phi_{12} \Phi_{22} \\ \Phi_{13} \Phi_{23} \\\Phi_{14} \Phi_{24} \\  \Phi_{11} \Phi_{22} +\Phi_{12} \Phi_{21} \\ \Phi_{11} \Phi_{23} +\Phi_{13} \Phi_{21}  \\  \Phi_{11} \Phi_{24} +\Phi_{14} \Phi_{21} \\ \Phi_{12} \Phi_{23} +\Phi_{13} \Phi_{22} \\ \Phi_{12} \Phi_{24} +\Phi_{14} \Phi_{22} \\ \Phi_{13} \Phi_{24} +\Phi_{14} \Phi_{23}   
},\quad
\calg_0(\theta):= \begmat{ \theta_1^2 \\ \theta_2^2 \\ \theta_3^2 \\ \theta_4^2 \\ \theta_1 \theta_2 \\ \theta_1\theta_3 \\ \theta_1 \theta_4 \\ \theta_2 \theta_3 \\ \theta_2 \theta_4 \\ \theta_3 \theta_4},
$$
where $\Phi_{ij}(t) \in \rea$ is the $(i,j)$-th element of the matrix $\Phi$.\\

We propose to estimate the parameters with the LS+DREM algorithm. Recall that this scheme---like all DREM-based estimators---generates {\it scalar} regression equations. Hence, we can restrict ourselves to the estimation of the first four parameters of $\calg(\theta)$. On the other hand, notice that $\calg(\theta) \in \rea^{14}$, hence a huge overparameterization was introduced by the procedure. 
\subsection{Surface-mount permanent magnet synchronous motor}
\lab{subsec44}
%
The classical, two-phase $\alpha\beta$ model of the non-salient PMSM is given by \cite{BOBetalaut15,KRAbook}
\begalis{
\dot x_1 &= -{R \over L}[x_1 +\lambda_m \cos(n_p x_3) +u_1]\\
\dot x_2 &=  -{R \over L}[x_2 +\lambda_m \cos(n_p x_3) +u_2]\\
\dot x_3 &= x_4\\
\dot x_4 &= -{f \over J} x_4 + {n_p \over J} (y_2 x_1 - y_1 x_2) -{\tau_L \over J},
}
$(x_1(t),x_2(t)) \in \rea^2$ is the flux vector, $x_3(t) \in [0,2\pi)$ is the rotor angle, $x_4(t) \in \rea$ the rotor angular velocity and $y(t) \in \rea^2$ is the {\it current}. The vector $u(t)\in \rea^2$ is the applied voltage, $R,f,J,n_p,L,\lambda_m$ are positive parameters and $\tau_L \in \rea$ is the {\it unknown} load torque, which is assumed constant. In the practically relevant scenario of {\it sensorless control} \cite{CHOetal}, the only measurable quantity is the {\it current}, which is related to the systems state via \cite[Equations (2) and (3)]{BOBetalaut15}:\footnote{Notice that the fist two equations of the systems model are, again, the consequence of Faraday's Law.}
\begequ
\lab{flucur}
\begmat{y_1 \\ y_2}={1 \over L}\begmat{x_1 \\ x_2}+{\lambda_m \over L} \begmat{\cos(n_p x_3) \\ \sin(n_p x_3)}.
\endequ

It is interesting to note that in this example there is  {\it no need} for the dynamic extension $\phi(x)$, consequently $z=x$. Thus, we select $n_z=4$ and choosing the mappings
\begalis{
W(y) =\begmat{0 & 0 & 0 & 0\\0 & 0 & 0 & 0\\ 0 & 0 & 0 & 1\\ {n_p \over J}y_2 & -{n_p \over J} y_1 & 0 & -{f \over J}},\;L(y,u)=\begmat{-R y_1 +u_1\\ -R y_2 +u_2\\0 \\ -{\tau_L \over J}},
}
yields the  system \eqref{dotz}. In contrast with the simplicity  of the derivations for the first step of the GPEBO design, the second step, namely the derivation of the regression equation is quite involved. Indeed, in this case the read out map $h(x,u)$ \eqref{flucur} has the trascendental functions $\sin(\cdot)$ and $\cos(\cdot)$ so, replacing directly in \eqref{flucur} the relation \eqref{x} will lead to a nonseparable regression equation of the form \eqref{nonsep}. On the other hand, it is possible to use the well-known fact that 
$$
\cos^2(n_p x_3)+\sin^2(n_p x_3)=1,
$$
to get an alternative algebraic equation
\begequ
\lab{algequ}
(Ly_1-x_1)^2+(Ly_2-x_2)^2=\lambda_m^2.
\endequ

We make at this point the important observation that, due to the particular form of the matrix $W$, the calculations are significantly simplified. Indeed, it is possible to show that for $j=1,2$, we have $\phi_j=\bfe_j$. Consequently, the first two elements of the vector $x$, defined in \eqref{x}, take the form
$$
x_j=\xi_j-\theta_j,\;j=1,2.
$$
Replacing this relation in \eqref{algequ}, and grouping terms, we obtain a separable NLPRE of the form \eqref{sepregegu} where 
\begalis{
\caly &:=(Ly_1-\xi_1)^2+(Ly_2-\xi_2)^2-\lambda_m \\
\psi & := \begmat{2(\xi_1-Ly_1) & \vdots & 2(\xi_2-Ly_2)& \vdots & 1 }\\  
\calg^\top(\theta)&:=\begmat{\theta_1 & \theta_2 &  \theta_1^2 + \theta_2^2}.
}
It is clear that the computation of the matrix $\Phi$ is {\it obviated} and the implementation of the GPEBO reduces to the calculation of 
\begequ
\lab{opeloo}
\dot \xi_j=-Ry_j+u_j,\;j=1,2.
\endequ

\begrem
\lab{rem5}
We underscore the fact that in the problem of sensorless motor control \cite{CHOetal} the main subject of interest is the estimation of the rotor flux, {\it e.g.}, $(x_1,x_2)$, from measurements of the electrical coordinates only---just as it is done in here. This is due to the fact that, knowing the flux, the rotor position can be easily reconstructed from the relation
$$
x_3={1 \over n_p}\arctan\Big({ Ly_2 - x_2 \over Ly_1-x_1}\Big),
$$
which is the standard procedure in all applications. The estimated value for the rotor angle becomes then
$$
\hat x_3={1 \over n_p}\arctan\Big({ Ly_2 - \xi_2 + \hat \theta_2 \over Ly_1-\xi_1+\hat \theta_1}\Big).
$$
\endrem

\begrem
\lab{rem6}
The PMSM observer presented here was first reported in \cite[Subsection 5.3]{ORTetalscl15} and is given here to show how it can be re-derived using the proposed immersion procedure. As discussed in that paper, the main drawback of this design is the implementation of an {\it open-loop integration}, {\it i.e.}, \eqref{opeloo}, that is sensitive to the presence of noise.     
\endrem

\begrem
\lab{rem7}
As indicated above, we can treat $\tau_L$, as well as $\lambda_m$,  as uncertain parameters an implement and adaptive GPEBO. 
\endrem
%
\subsection{Mechanical systems}
\lab{subsec45}
%
Consider an $n_q$-degrees of freedom (DoF) mechanical system expressed in the coordinates position $q(t) \in \rea^{n_q}$ and momenta $p(t) \in \rea^{n_q}$, with output $y=q$. The dynamics takes the form
\begalis{
\dot q &=M^{-1}(y)p\\
\dot p &= \hal \nabla_y[p^\top M^{-1}(y)p]-\nabla V(y)+G(y)u,
}  
where $M:\rea^{n_q} \to \rea^{n_q \times n_q}$ is the positive definite systems inertia, $V:\rea^{n_q} \to \rea$ is the potential energy, $G:\rea^{n_q} \to \rea^{n_q \times m}$ is the input matrix and $u(t) \in \rea^{m}$ are the input signals. 

From the second equation above it is clear that the viability of the GPEBO design method of Proposition \ref{pro1} is uniquely determined by the ``form" of the vector $\nabla_y[p^\top M^{-1}(y)p]$. Although a general answer to this question is not easy to establish we present here two practically relevant examples where Proposition \ref{pro1} is applicable. \\

\begin{figure}[htp!]
\centering
\includegraphics[width=0.3\linewidth]{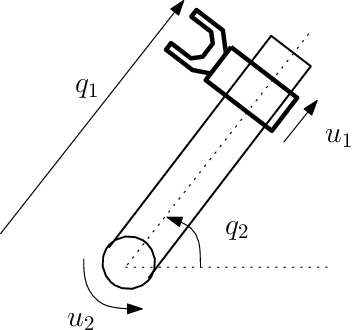}
\caption{Two DoF prismatic robot.}
\label{fig1}
\end{figure}

\noindent{\bf First example:} Consider the two DoF prismatic robot example \cite{ANGSONWAN} depicted in Fig. \ref{fig1}.
Denoting $x=\col(q_1,q_2,p_1,p_2)$ the inertia and input matrices are of the form 
$$
M(x_2)=\begmat{a x_2^2+b & 0 \\ 0 & a}, \quad G=I_2,
$$
where $a,b$ are positive parameters. There is no potential energy term and the dynamics is given as
\begalis{
\dot x_1 &= {1 \over a x_2^2+b}x_3,\;\dot x_2 = {1 \over a}x_4\\
\dot x_3 &= u_1,\;\dot x_4 = -{2ax_2 \over (a x_2^2+b)^2}x^2_3+u_2\\
y&=\begmat{x_1 \\ x_2}.
}
Selecting $n_z=5$ and choosing the mappings
\begalis{
\phi(x) &=\hal x_3^2,\;W(y,u) =\begmat{0 & 0 & {1 \over a y_2^2+b} & 0 & 0\\ 0 & 0 & 0  & {1 \over a}& 0\\ 0 & 0 & 0 & 0 & 0\\  0 & 0 & 0 & 0 &-{4ay_2 \over (a y_2^2+b)^2}\\ 0 & 0 & u_1 & 0 & 0},\;L(u)=\begmat{0 \\ 0 \\ u_1\\ u_2 \\0},
}
yields the system \eqref{dotz}. The corresponding GPEBO is given as
\begalis{
\dot \xi&=W(y,u)\xi+L(u)\\
\dot \Phi&=W(y,u) \Phi,\;\Phi(0)=I_{5}
}
and the LRE is \eqref{lreypsi} with
$$
\caly:=\begmat{I_2 & \vdots & 0_{2 \times 3}}\xi-y,\; \psi:=\begmat{I_2 & \vdots & 0_{2 \times 3}}\Phi.
$$

\begin{figure}[htp!]
\centering
\includegraphics[width=0.4\linewidth]{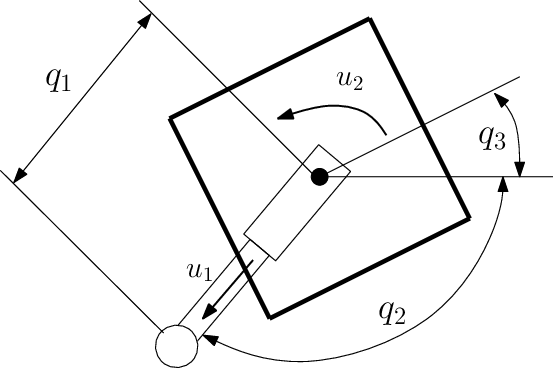}\\
\caption{Robotic leg}
\label{fig2}
\end{figure}

\noindent{\bf Second example:} Consider the robotic leg, which was studied in \cite{BULLEWbook,VENetal} and is depicted in Fig. \ref{fig2}. It has three DoF and two control forces $u=\col(u_1, u_2)$. Thus, denoting $x=\col(q_1,q_2,q_3, p_1, p_2, p_3)$, its inertia and input matrices are of the form 
$$
M(x_1)=\begmat{ m_1 & 0 & 0\\ 0 & m_1 x_1^2 & 0 \\ 0 & 0 & m_2}, \quad G=\begmat{1 & 0 \\ 0 & 1\\ 0 & -1 },
$$
with positive numbers $m_1$ and $m_2$. There is no potential energy term and the dynamics, assuming the leg length is measurable, is given as 
\begalis{
\dot x_1 &= {1 \over m_1}x_4, \quad  \dot x_2 = {1 \over m_1 x_1^2}x_5\\
\dot x_3 &= {1 \over m_2}x_6, \quad \dot x_4 = {1 \over m_1 x_1^3}x^2_5 +u_1\\
\dot x_5 &= u_2, \quad \quad \dot x_6 = -u_2\\
y&=x_1.
}
Selecting $n_z=7$ and choosing the mappings
\begalis{
\phi(x) &=x_5^2,\;W(y,u) =\begmat{0 & 0 & 0 &{1 \over m_1} & 0 & 0 & 0\\ 0 & 0 & 0  & 0 & {1 \over m_1 y^2}& 0 & 0\\ 0 & 0 & 0 & 0 & 0 & {1 \over m_2} & 0\\   0 & 0 & 0  & 0 & 0& 0 &  {1 \over m_1 y^3}\\ 0 & 0 & 0 & 0 & 0 & 0 &0\\  0 & 0 & 0 & 0 &0 & 0& 0 \\ 0 & 0 & 0 & 0 & 2u_2 & 0 & 0},\;L(u)=\begmat{0 \\ 0 \\ 0 \\ u_1 \\ u_2 \\ -u_2 \\ 0},
}
yields the the system \eqref{dotz} and the design of the LRE and the GPEBO proceeds as done in the example above.

\begrem
\lab{rem8}
A class of mechanical systems for which GPEBO applies directly are the so-called partially linearizable via coordinate change. As the name indicates this class can be transformed to a system {\it linearly dependent} on momenta via a change of coordinates in the momenta and it has been geometrically characterized in \cite{VENetal}---see also \cite{CHAMCL}. This class contains several practically relevant examples, including the cart-pendulum system, the 3-link planar manipulator and the planar redundant manipulator with one elastic DoF. 
\endrem
\section{Simulations}
\lab{sec5}
To illustrate the performance of the proposed observers, in this section we present simulations of the magnetic levitation system and the prismatic robot example using the LS+DREM estimator\cite{ORTROMARA}, which is summarized in Appendix \ref{appa}.
 \subsection{Magnetic levitation system}
\begin{figure}[htp!]
\centering
\includegraphics[width=0.6\linewidth]{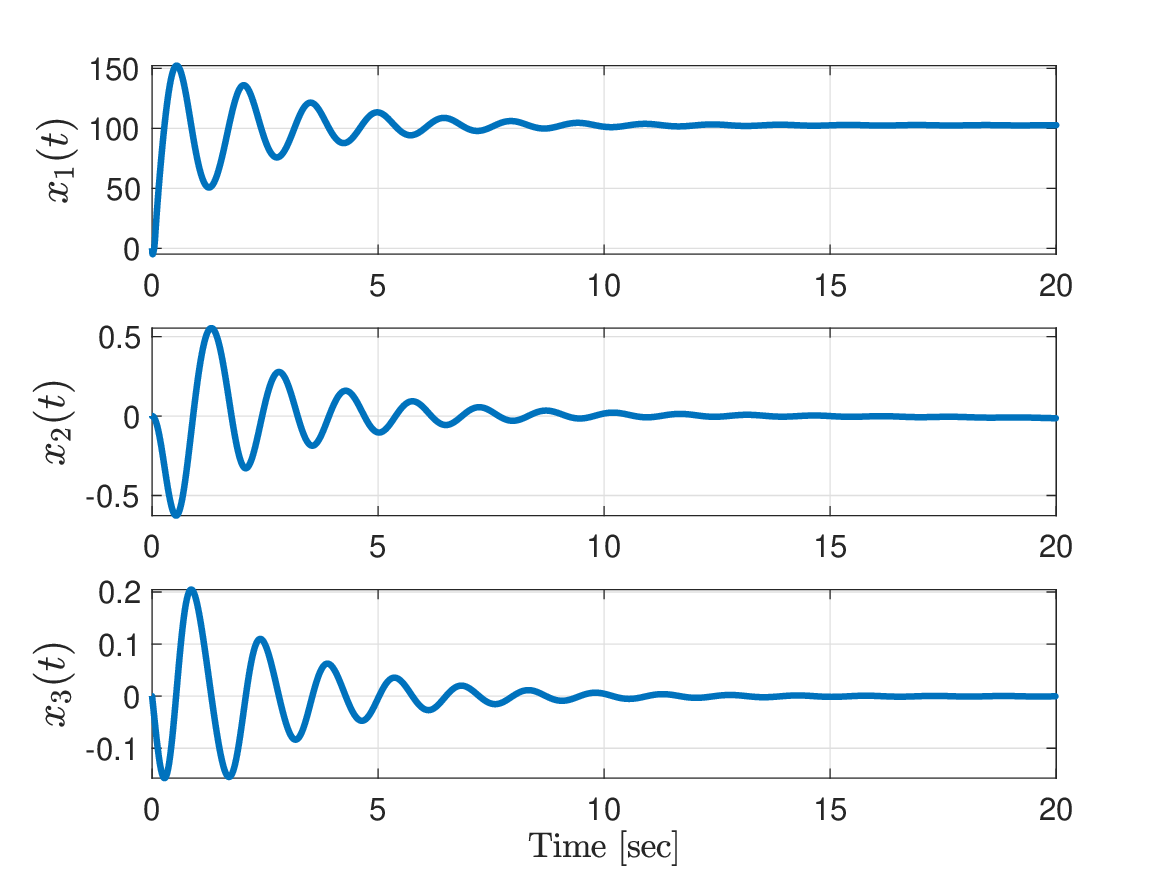}\\
\caption{Transient behavior of the state $x(t)$}
\label{figx}
\end{figure}

\begin{figure}[htp!]
\centering
\includegraphics[width=0.6\linewidth]{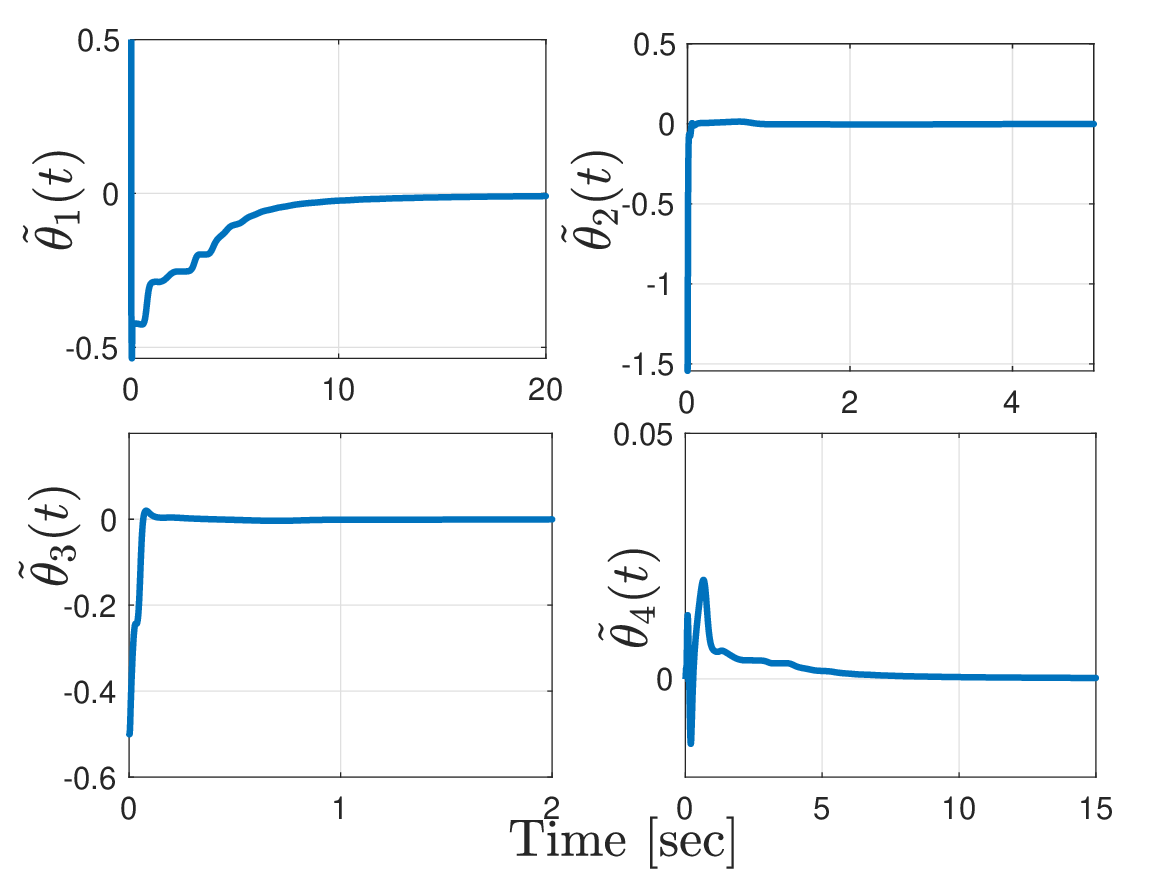}\\
\caption{Transient behavior of the parameter estimation errors $\tilde \theta(t)$}
\label{figeth}
\end{figure}

\begin{figure}[htp!]
\centering
\includegraphics[width=0.6\linewidth]{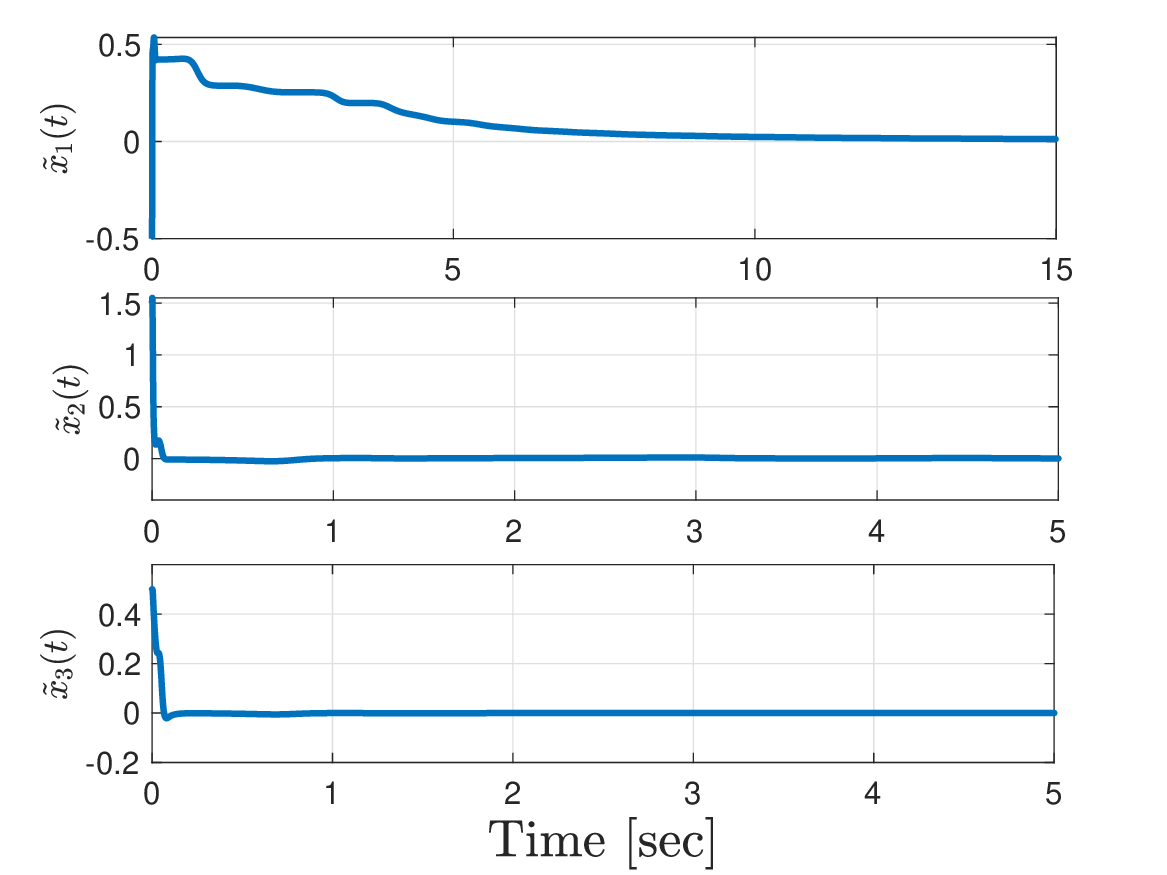}\\
\caption{Transient behavior of the state estimation errors $\tilde x(t)$}
\label{figex}
\end{figure}

 \lab{subsec51}
 The simulations for this example were carried out using a {\it certainty equivalence} version of the controller given in  \cite{ORTelat01}. Namely,
 $$
 u=\frac{R}{k} y -K_p\left( \frac{1}{\alpha} ( \hat x_1 -x_{1_\star}) +(\hat x_2 - x_{2_\star})\right)-\left( \frac{\alpha}{m} +K_p \right)\hat x_3
 $$
where $x_{1_\star}=\sqrt{2kmg}$ and $x_{2_\star}=0.01$ are the the desired values of $x_1$ and $x_2$, respectively. The  plant parameters were taken from \cite{YIetal} and are given by  $m=0.0844 kg$, $R=2.52 \Omega$, $g=9.81m/s^2$, $k=6404 \mu H m$. The control gains were set as $K_p=400$ and $\alpha=80$. For the LS+DREM estimator the following numerical values were used in the simulation. We set the initial conditions $x(0)=\col(0,0,0)$ and $\xi(0)=\col(1,3,2,1.5)$. Consequently, the unknown $\theta$ is given by $\theta= \xi(0)- x(0)=\col(1,3,2,1.5)$. In particular, in this example, we only utilize the LS part of the estimator with  all initial conditions of $\calh_0$ equal to 1.5. The gains of the LS estimator were picked as $\gamma_\calh=680$, $\chi_0 = 150$ and $f_0=10$.  We solely show the first four elements of $\hat \calh$ that correspond to the estimation of $\theta$, which are the parameters of interest here. In Fig. \ref{figx} we appreciate the transient behavior of the system state $x$. Fig. \ref{figeth} shows the excellent behavior of the parameter estimation errors $\tilde \theta:=\hat \theta -\theta$. Finally, Fig. \ref{figex} shows the transient behavior of the state estimation errors  $\tilde x:=\hat x - x$.

 \subsection{Prismatic robot}
 \lab{subsec52}
 The simulations for this example are conducted using a simple certainty equivalent PD controller, which has the  form 
$$
u= - K_p \begmat{x_{1}- x_{1_\star}\\ x_{2}- x_{2_\star}} - K_d\begmat{\hat x_{3}\\ \hat x_{4}},   
$$ 
 with the desired point $ (x_{1_\star},x_{2_\star})= \left(0.1, \frac{\pi}{8}\right)$ and control gains $K_p= \diag\{70, 30 \}$ and $K_d= 10 I_2$.
 
 The  plant parameters  are given by  $a=1 Kg$ and $b=3 Kg m^2$ and they were taken from \cite{ROMDONORT}. For the LS+DREM  estimator the following numerical values were used in the simulation. We set the initial conditions $x(0)=\col(0,0,0,0)$ and $\xi(0)=\col(1,2,3,4,5)$. Consequently, the unknown $\theta$ is given as $\theta= \xi(0)- x(0)=\col(1,2,3,4,5)$. The initial conditions of the estimator were set to $\calh_0 =\col(1, 1, 1, 1, 1)$,  the gains of the LS part  were picked as $\gamma_\calh=1.8$, $\chi_0 = 5$ and $f_0=20$ and the gain of the scalar gradient estimators were chosen as $\gamma_i=1100$ for $i=1..5$
 
\begin{figure}[htp!]
\centering
\includegraphics[width=0.6\linewidth]{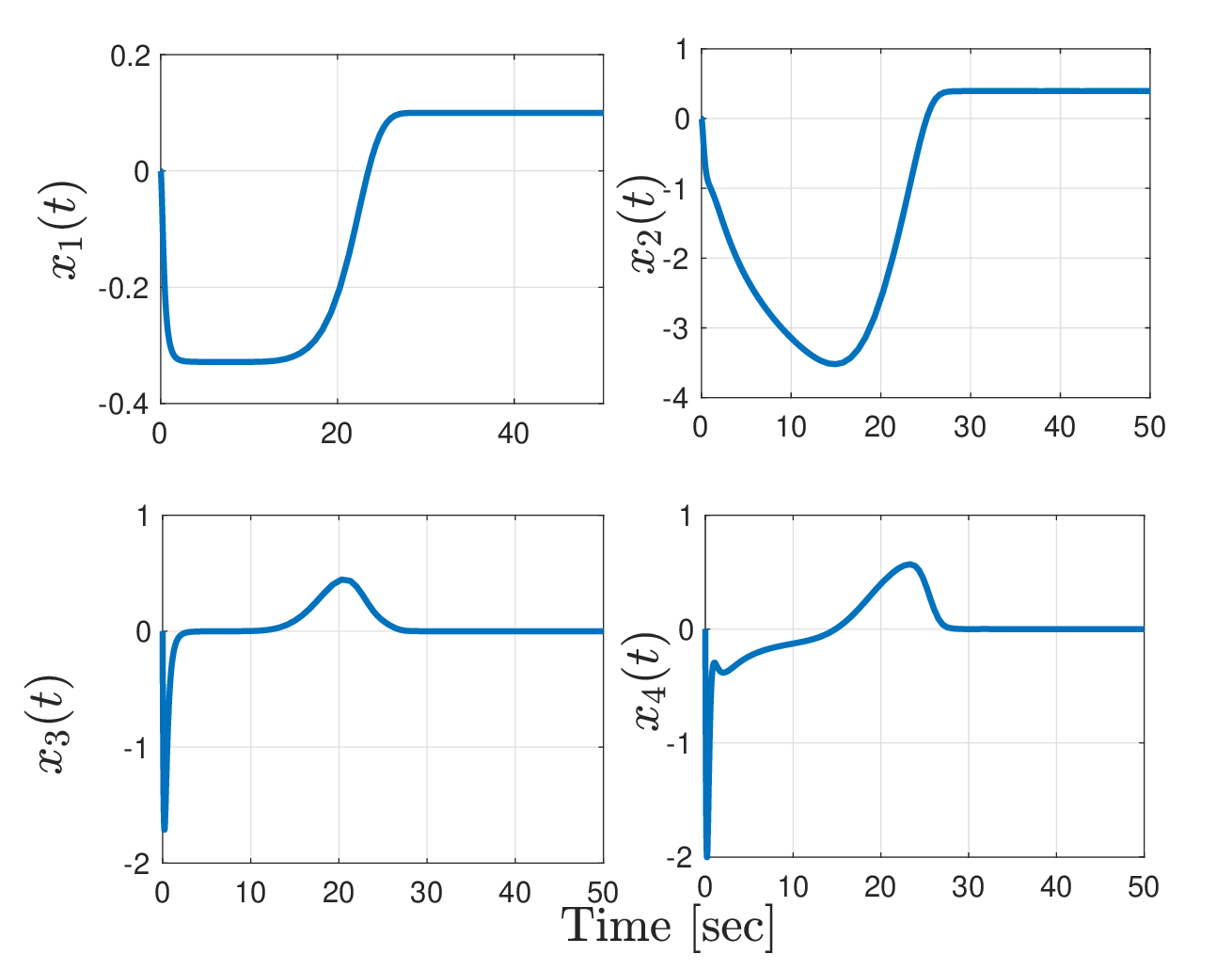}\\
\caption{Transient behavior of the state $x(t)$}
\label{figxp}
\end{figure}

\begin{figure}[htp!]
\centering
\includegraphics[width=0.6\linewidth]{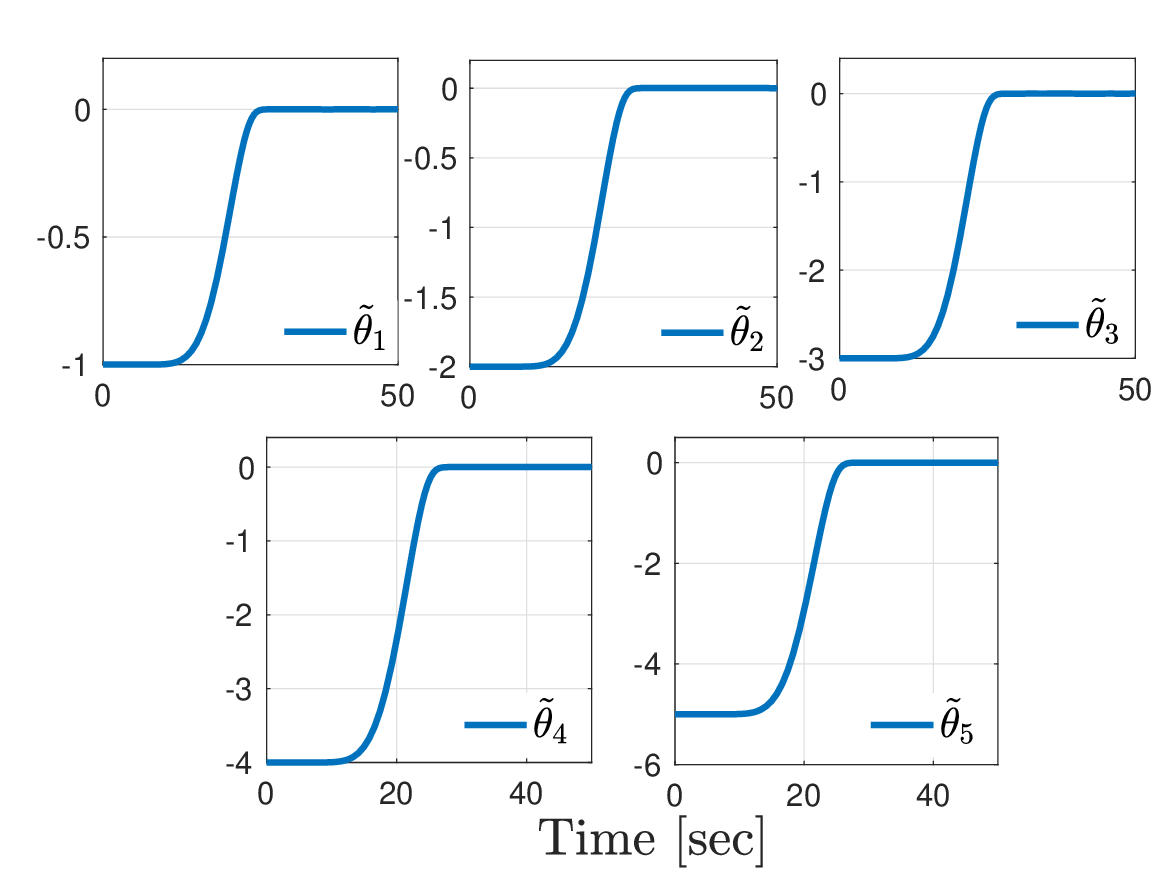}\\
\caption{Transient behavior of the parameter estimation errors $\tilde \theta(t)$}
\label{figethp}
\end{figure}

\begin{figure}[htp!]
\centering
\includegraphics[width=0.6\linewidth]{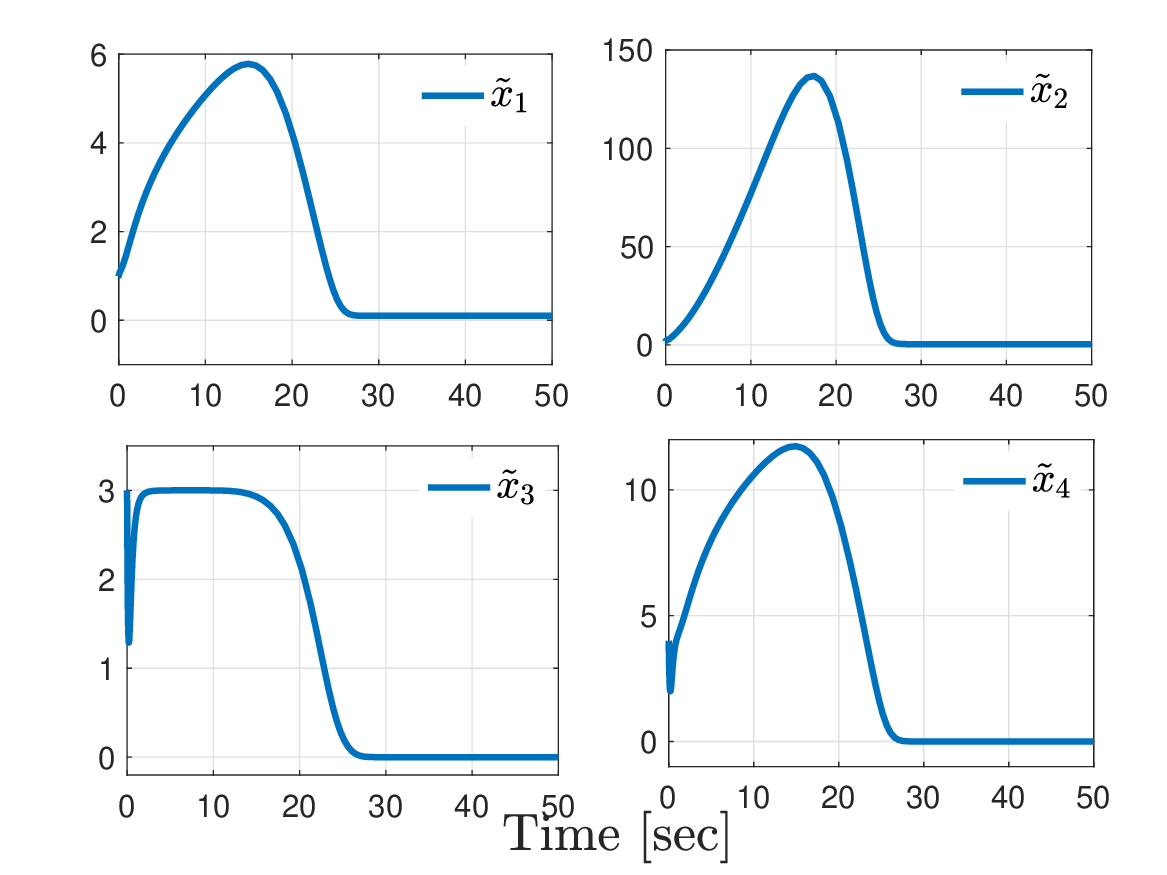}\\
\caption{Transient behavior of the state estimation errors $\tilde x(t)$}
\label{figexp}
\end{figure}

  In Fig. \ref{figxp} we appreciate the transient behavior of the system state
$x$. Fig. \ref{figethp} shows the excellent behavior of the parameter estimation errors $\tilde \theta$. Finally,  Fig. \ref{figexp} shows the transient behavior of the state estimation errors  $\tilde x$.

%
\section{Concluding Remarks and Future Research}
\lab{sec6}
%
We have presented in the paper an algebraic procedure to immerse a general nonlinear system into a larger dimensional state-affine one in order to design a GPEBO. Although the procedure relies on a case-by-case example, we show that it is rather systematic and useful to solve several interesting examples. In particular, it has revealed very interesting novel parameterizations for physical examples that have been extensively studied. In particular the PMSM system is a benchmark in observer and output feedback control problem that has attracted the attention of many research groups\cite{BERPRA,VERetal} and references therein. 

Although, in principle, it could have been possible to consider a transformation of the first $n$ states of the vector $z$, that is, to define $z=\col(\phi_0(x),\phi(x))$, we have opted in the paper to avoid this generalization. This was done in order to easily obtain the relation \eqref{x}, which is the first step in GPEBO design. Otherwise, it would be necessary to ensure the injectivity of the mapping $\phi_0$.  

Our current research efforts are precisely to explore the possibilities of exploiting the new parameterizations of the physical examples and compare their performance with the many existing solutions to the observer design problems. Also, motivated by the developments of \cite[Subsection 6.3.4]{BERbook}, we are currently investigating the use of {\it time-varying} transformations.
%

\section*{Disclosure Statement}

We have no conflicts of interest to disclose.


%
\appendix
\section{The Least Squares  plus Dynamic Regression Extension and Mixing Parameter Estimator}
\label{appa}

The main properties of the LS+DREM estimator are summarized in the proposition below, whose proof may be found in \cite{ORTROMARA}.\\

\begpro
\lab{pro3}
Consider the regression equation \eqref{sepregegu} and assume $\psi(t)$  is IE and {\it bounded}.  Define the LS+DREM estimator with forgetting factor \cite{ORTROMARA}.
\begsubequ
\lab{lsd}
\begali{
\lab{lsd1}
		\dot{\hat \calg} & =\gamma_\calg F \psi^\top (\caly-\psi \hat\calg),\; \hat\calg(0)=\calg_{0} \in \rea^{n_\psi}\\
\lab{lsd2}
\dot {F}& =  -\gamma_\calh F \psi^\top \psi F +\chi F,\; F(0)={1 \over f_0} I_{n_\psi} \\
\lab{lsd3}
\dot{\hat \theta}_i & = \gamma_i  \Delta [{\bf Y}_i -\Delta \hat\theta_i ],\; \hat\theta_i(0)=\theta_{i0} \in \rea, \\
\dot z &=\; -\chi z, \; z(0)=1, \\
\chi &= \chi_0 \left( 1-{{\| F\|}\over{k}} \right)
}
\endsubequ
where $i=1,\dots,n_z$,	with tuning gains the scalars $\gamma_\calh>0, \gamma_i>0$, $f_0>0$, $\chi_0> 0$ and $k\geq \frac{1}{f_0}$; and we defined the signals
	\begalis{
		\Delta & :=\det\{I_{n_\psi}-z f_0F\}\\
		{\bf Y} & := \adj\{I_{n_\psi}- zf_0F\} (\hat\calg -  zf_0F \calg_{0}),
	}
where $ \adj\{\cdot\}$ denotes the adjugate matrix.
\begenu
\item [{\bf (i)}] For all initial conditions the estimated parameters verify
$$
\liminf|\hat \theta(t)- \theta|=0,
$$
exponentially fast.
\item [{\bf (ii)}] The state estimation 
\begequ
\lab{hatx}
\hat x= D(\xi - \Phi \hat \theta)
\endequ 
verifies \begali{
&\lim_{t\to\infty} |\hat x(t)-x(t)|=0,
\lab{concon}
}
exponentially fast.
\item [{\bf (iii)}] All the signals are {\it bounded}.
\endenu 
\endpro

\end{document}